\documentclass[letterpaper]{article} 
\usepackage{aaai24}  
\usepackage{times}  
\usepackage{helvet}  
\usepackage{courier}  
\usepackage[hyphens]{url}  
\usepackage{graphicx} 
\usepackage{natbib}  
\usepackage{caption} 
\usepackage{algorithm}
\usepackage{algorithmic}

\usepackage{newfloat}
\usepackage{listings}
\DeclareCaptionStyle{ruled}{labelfont=normalfont,labelsep=colon,strut=off} 
\floatstyle{ruled}
\newfloat{listing}{tb}{lst}{}
\floatname{listing}{Listing}
\usepackage{nameref}

\usepackage{chngcntr}

\title{The Impact of Responsible AI Research on Innovation and Development}
\author {
    Ali Akbar Septiandri\textsuperscript{\rm 1},
    Marios Constantinides\textsuperscript{\rm 1},
    Daniele Quercia\textsuperscript{\rm 1, 2}
}
\affiliations {
    \textsuperscript{\rm 1}Nokia Bell Labs, Cambridge, UK\\
    \textsuperscript{\rm 2}Kings College London, London, UK\\
    \{ali.septiandri, marios.constantinides, daniele.quercia\}@nokia-bell-labs.com
}

\usepackage{color, colortbl}
\usepackage{enumitem}
\usepackage{booktabs}

\begin{document}

\maketitle

\begin{abstract}
Translational research, especially in the fast-evolving field of Artificial Intelligence (AI), is key to converting scientific findings into practical innovations. In Responsible AI (RAI) research, translational impact is often viewed through various pathways, including research papers, blogs, news articles, and the drafting of forthcoming AI legislation (e.g., the EU AI Act). However, the real-world impact of RAI research remains an underexplored area. Our study aims to capture it through two pathways: \emph{patents} and \emph{code repositories}, both of which provide a rich and structured source of data. Using a dataset of 200,000 papers from 1980 to 2022 in AI and related fields, including Computer Vision, Natural Language Processing, and Human-Computer Interaction, we developed a Sentence-Transformers Deep Learning framework to identify RAI papers. This framework calculates the semantic similarity between paper abstracts and a set of RAI keywords, which are derived from the NIST's AI Risk Management Framework; a framework that aims to enhance trustworthiness considerations in the design, development, use, and evaluation of AI products, services, and systems. We identified 1,747 RAI papers published in top venues such as CHI, CSCW, NeurIPS, FAccT, and AIES between 2015 and 2022. By analyzing these papers, we found that a small subset that goes into patents or repositories is highly cited, with the translational process taking between 1 year for repositories and up to 8 years for patents. Interestingly, impactful RAI research is not limited to top U.S. institutions, but significant contributions come from European and Asian institutions. Finally, the multidisciplinary nature of RAI papers, often incorporating knowledge from diverse fields of expertise, was evident as these papers tend to build on unconventional combinations of prior knowledge.
\end{abstract}

\section{Introduction}
\label{sec:introduction}

Translational research plays a critical role in bridging the gap between scientific discoveries and practical applications, a process vital for technological innovation and societal advancement. This is especially pertinent in the field of Artificial Intelligence (AI), where theoretical advancements can have profound implications. For example, \citet{vieira2021understanding} discussed how machine translation, a subset of AI, contributes to societal and technological innovation by applying AI in language translation for medical and legal use cases. \citet{mayrink2022translational} emphasized that translational research is key to accelerating the innovation process, transforming basic science into applied science and innovation. \citet{horowitz2017accelerators} highlighted how translational research, through ``accelerators,'' can rapidly generate novel solutions to eliminate disparities in healthcare, and \citet{eweje2022translatability} discussed its potential in biomedical research for drug discovery. In Human-Computer Interaction (HCI), \citet{cao2023breaking} demonstrated that HCI research significantly influences systems-oriented research through an analysis of 70 thousand patent citations. Similarly, \citet{ahmadpoor2017dual} found that fields such as nanotechnology and computer science are closely linked to patents, providing evidence of the direct impact of research on technological advancement. 

The research problem of studying the multi-faceted impact of Responsible AI (RAI) has been long recognized. Indeed, a prominent 2022 FAccT paper called for more research on real-world impact, including the transfer of research into repositories and software products~\cite{laufer2022four}. It is also equally important to recognize the translational impact of RAI research through frameworks, policies, and governance mechanisms that guide AI ethical deployment. This demonstrates the diverse pathways through which RAI research can influence societal practices and regulations (e.g., EU AI Act~\cite{eu_2021aiAct}, NIST AI Risk Management Framework~\cite{nist_framework}). Nevertheless, its influence on patents and code repositories has not been fully explored. As AI becomes increasingly integrated into various sectors, it is vital to consider the ethical, fair, and sustainable development of AI technologies~\cite{tahaei2023human}, and how these technologies impact the society through technological innovations and development.

To address this gap, we aim to identify the factors that contribute to the impact of RAI.\footnote{Project website: \url{https://social-dynamics.net/rai-impact}} Our goal is to understand two pathways (i.e., through patents and code repositories) through which RAI research can drive innovation and development. In so doing, we made two main contributions:
\begin{enumerate}
    \item We obtained a large dataset of 200K research papers from 1980 to 2022 in the fields of AI, Computer Vision, Data Mining, HCI, and NLP. This dataset includes 85K citations linking these papers to patents from the United States Patent and Trademark Office (USPTO), and 1 million GitHub repositories associated with these papers. Using a Sentence-Transformers Deep Learning framework for NLP and the NIST framework keywords, we identified 1,747 RAI papers that served as the basis for our analysis. We developed a set of four metrics that capture the amount of research impact on patents and repositories, its time evolution, and the main factors that shape it, including authors' institutions and papers' conventional combination of prior knowledge.
    \item We found that a small subset of RAI papers that go into patents or repositories is highly cited, with this process taking between 1 year for repositories up to 8 years for patents. When examining the top institutions producing impactful RAI papers in terms of patents and repositories, leading U.S. institutions are not the only contributors; significant input also comes from European and Asian institutions, reflecting a broader global influence. Finally, RAI papers typically incorporate unconventional knowledge, merging insights from diverse fields such as Machine Learning (ML) and HCI. 
\end{enumerate}

In light of these results, we discuss our study's implications for measuring RAI impact and encouraging translational research.
\section{Related Work}
\label{sec:related_work}
AI advances, particularly in specialized tasks such as language generation, are deeply intertwined with the impact of research on innovation and development. Next, we discuss literature on RAI research's impact, and more broadly, the scientific impact on innovation and development.

\subsection{Responsible AI Research Impact}
\label{subsec:beyond}
Abebe et al.~\cite{abebe2020roles} argued that technical work can drive social change as a: 1) diagnostic tool to accurately measure social issues; 2) formalizer to define problems and explore possible solutions; 3) rebuttal to delineate the limits of what technical solutions can achieve; and 4) synecdoche by bringing public awareness to social challenges. However, as Laufer et al.~\cite{laufer2022four} caution, the journey towards leveraging RAI scholarship for positive change may be hindered by economic or political interests. These interests might suppress important ethical discussions (not everything that can be built ought to be), and suggest that the best solution to a problem may not necessarily be a technical one~\cite{selbst2019fairness} but may well require organizational cultural shift~\cite{rakova2021responsible, balakrishnan2011research}. 

RAI research extends its impact through various channels by raising awareness through blogs, news articles, and research papers that may not go into patents or repositories. For example, \citeauthor{laufer2022four} conducted a meta (reflexive) study on four years of FAccT proceedings to extract research topics (e.g., group-level fairness and disinformation) and understand community's values (e.g., industry influence over published research). \citet{septiandri2023weird} analyzed 128 papers published at FAccT between 2018 and 2022, and found that 84\% of these papers were exclusively based on Western countries' participants, particularly from the U.S. (63\%). 

This points to difficulties and challenges associated with ensuring data inclusivity, mainly stemming from biases in the data collection process~\cite{olteanu2019social, baeza2018bias, madaio2022assessing}. That is why conferences like the  Neural Processing Information Systems (NeurIPS)~\cite{ashurst2020guide} and the International Conference on Machine Learning (ICML) have already started mandating statements including ``any risks associated with the proposed methods, methodology, application or data collection and data usage''. More recently, \citet{olteanu2023responsible} argued that RAI research needs impact statements, too; such statements aim at disclosing any possible negative consequences, contributing to more inclusive research. Additionally, researchers have proposed algorithmic impact assessments as a form of accountability for organizations that build and deploy automated decision-support systems~\cite{metcalf2021algorithmic}, and frameworks for bridging the gap between technical and ethical aspects of AI systems~\cite{kasirzadeh2021reasons}.

RAI research draws from concepts such as responsible research and innovation. Brundage~\cite{brundage2016artificial} argued that the design of AI should balance social impacts, reflect on theory, and involve public dialogue, while Owen et al.~\cite{owen2012responsible} emphasized innovations’ unpredictable consequences. To facilitate responsible research and innovation, Stilgoe et al.~\cite{stilgoe2020developing} proposed a framework for anticipating, reflecting, engaging, and acting upon challenges such as the narrow focus on technological innovation~\cite{blok2015emerging}.
RAI can significantly influence policy development. Documentation practices of datasheets for datasets~\cite{gebru2021datasheets} and model cards~\cite{mitchell2019model} are now referenced in the draft of the EU AI Act~\cite{eu_2021aiAct}, while the Ethics guidelines for Trustworthy AI~\cite{eu_trustworthyAI} proposed by EU High-level group on Trustworthy AI based the definition of trust in scientific findings~\cite{siau2018building}. \citet{naidoo2022artificial} emphasized the need for a national policy framework in South Africa addressing outdated legislation, data and algorithmic bias, and liability dilemmas for responsible development and deployment. \citet{onder2021roles} demonstrated how AI could contribute to crisis management policies by aiding in various stages like preparation, mitigation-prevention, response, and recovery, providing decision support for high-quality decisions. RAI research has also been used for community engagement. \citet{verdiesen2018design} highlighted the need for guidelines and regulations in AI design and development, considering its ethical, legal, and societal impact. \citet{deshpande2022responsible} identified various stakeholders in AI systems, including individual, organizational, and international stakeholders, emphasizing the diverse roles and responsibilities in developing ethical AI systems.
\smallskip

\subsection{Measuring Research Impact on Innovation}
\label{subsec:innovation}
Research impact on innovation is frequently gauged through the extent to which scientific knowledge transitions into products and services~\cite{jefferson2018mapping, cao2023breaking, barnes2015use, mendoza2018systematic, brooks1994relationship}. \citet{scarra2022impact} identified several mechanisms that drive this impact such as technology transfer and collaboration, illustrating the multifaceted ways in which research influences innovation~\cite{ahmadpoor2017dual}.
 
A widely used source to measure research impact is through patents as they are a typical source to identify emerging technological innovations~\cite{straccamore2023urban, straccamore2023geography, arcaute2015constructing}.
\citet{cao2023breaking} analyzed 70,000 patent citations from leading HCI research venues, revealing a significant impact of HCI in systems-oriented research. \citet{tijssen2001global} investigated the influence of Dutch-authored research papers on global inventions. While their methodology is similar to our approach, they mainly used simple bibliometric counting, which might not fully reflect these indirect influences. The field of scientometrics, employing network science techniques, offers a quantitative lens to explore scientific research's dynamics and influence~\cite{fortunato2018science, sinha2015overview}. By analyzing 4.8 million U.S. patents and 32 million research papers, \citet{ahmadpoor2017dual} found that mathematics was the most remote field from patents, whereas nanotechnology and computer science were among the closest. \citet{mariani2019early} examined the U.S. patent citation network from 1926 to 2010 and introduced a rescaled PageRank metric, which outperformed traditional citation counts in identifying historically significant patents early. In a similar vein, \citet{park2023papers} studied data over six decades, encompassing 45 million papers and 3.9 million U.S. patents, and observed that recent papers and patents tend to adhere more to established science and technology trends, reducing their likelihood of deviating from traditional field norms to initiate new research directions.

\subsection{Measuring Research Impact on Development}
\label{subsec:development}
The impact of research on software development and code repositories, while less explored, presents an interesting dimension of technological advancement~\cite{badashian2016measuring, badashian2014involvement, blincoe2016understanding}. \citet{alqahtani2016tracing} demonstrated that traceability links in software repositories enhance knowledge sharing, highlighting the importance of cross-repository knowledge transfer for improving software security. \citet{inokuchi2019academia} explored the role of papers as knowledge sources for software development, particularly in algorithm knowledge transfer, highlighting the direct impact of academic research on software development practices.

A significant body of literature has explored the characteristics of influential GitHub repositories. 
These repositories often possess distinct features that elevate their visibility and impact within the software development community. For example, they are characterized not only by high numbers of followers but also by low number of coding violations~\cite{diamantopoulos2020employing}. Furthermore, well-maintained issue pages with frequent label usage and a higher number of issues are also indicators of influential repositories~\cite{yamamoto2020metrics}. In addition to these factors, \citet{safari2020analysis} found that the most influential users and projects on GitLab are associated with the founding team, indicating that the influence may also be attributed to the prominence of the contributors. \citet{borges2016predicting} highlighted that repositories with a high number of stars are influential and that their popularity can be predicted based on growth trends and similarities with other repositories.
\smallskip

\noindent\textbf{Research Gaps.}
As AI continues to advance, little is known about whether the scientific efforts to ensure that AI is fair, transparent, and accountable have made a translational impact. To address this gap, this study aims to measure the impact of RAI research on both innovation and development. As previously discussed, the impact may extend beyond innovation and development through blogs and news articles. However, tracking the impact on these channels requires structured data, which are currently unavailable (compared to publicly available data on patents and code repositories).
\section{Methodology}
\label{sec:method}

Drawing upon previous literature~\cite{cao2023breaking, ahmadpoor2017dual, manjunath2021comprehensive, uzzi2013atypical}, we set out to understand the factors that make these RAI papers impactful, including temporal aspects, authors' affiliations, and the extent to which the papers draw upon (un)conventional knowledge. More specifically, we investigated four guiding questions:

\begin{itemize}[leftmargin=7mm]
     \item[\textbf{Q\textsubscript{1}}:] \textit{Which types of RAI papers go into patents and repositories?}
    \item[\textbf{Q\textsubscript{2}:}] \textit{How long does it take for RAI papers to go into patents or repositories?}
    \item[\textbf{Q\textsubscript{3}:}] \textit{Which institutions produce RAI papers that go into patents or repositories?}
    \item[\textbf{Q\textsubscript{4}:}] \textit{How does combining conventional and unconventional knowledge in RAI papers affect their impact?}
\end{itemize}

These questions differ from traditional research questions that involve hypotheses; instead, they focus on identifying factors that contribute to the impact of RAI papers. They involve significant work in linking multiple datasets in ways that have not been done before on such a large scale. To this end, we collected, analyzed, and make publicly available a large-scale dataset focusing on three elements: \emph{papers}, \emph{patents}, and \emph{repositories}. Next, we explain the data collection and processing steps to identify RAI papers, state our method's limitations, and define metrics for our analysis.

\subsection{Data Collection}
\label{subsec:data_collection}
\noindent{\textbf{Papers.}}
From the May 2023 release of Semantic Scholar,\footnote{https://www.semanticscholar.org/product/api} we collected 200 million papers and retained papers based on three criteria: \emph{1)} the presence of detailed metadata fields such as `venue', `title', `year', `citation count', `citations', `DOI', `ArXiv id', `authors', `is open access' and `abstract'; \emph{2)} the publication timeframe to be between 1980 and 2022; and \emph{3)} the papers to be written in English. Through this process, we compiled a dataset of 50 million papers.

\smallskip
\noindent{\textbf{Patents.}}
From Google Patents Public Dataset, we collected 15 million patents filed in the United States Patent and Trademark Office (USPTO) between 1980 and 2022. These patents are characterized by various attributes, including the publication number, country code, date of publication, data about the inventors, the abstract, title, and references to non-patent literature. We extracted these attributes using the BigQuery service.\footnote{https://console.cloud.google.com/marketplace}

\smallskip
\noindent{\textbf{Repositories.}}
From ``Papers With Code'',\footnote{https://paperswithcode.com/api/v1/docs} we collected 1 million papers that had at least one link to a GitHub repository. This resulted in a set of papers from the inception of GitHub in 2008 through to 2022.

\subsection{Data Processing}
\label{subsec:data_preprocessing}
\begin{figure*}
    \centering
    \includegraphics[width=1\textwidth]{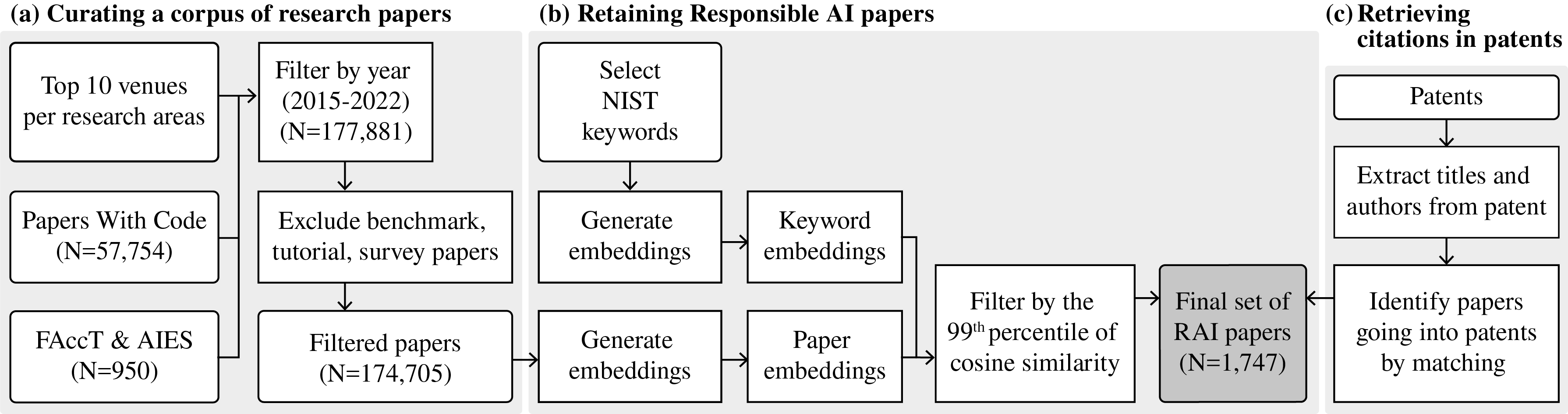}
    \caption{Data processing overview. \emph{a)} curated papers from three sources, filtered them by year, and excluded non-research-oriented papers; \emph{b)} identified RAI papers by comparing the embeddings (generated using ST5~\cite{ni2021sentence}) of paper abstracts and selected NIST keywords; \emph{c)} retrieved patent citations by matching the titles of patent references and those of RAI papers.}
    \label{fig:schema}
\end{figure*}

We followed three steps (Figure~\ref{fig:schema}). First, we curated a set of papers from a list of selected research areas and papers with code (Table A.5, Supplementary Information). Second, we retained those papers that are about RAI using a keyword-based approach based on the NIST Framework. Third, we extracted citations by matching papers referenced in patents with the Semantic Scholar metadata.

\smallskip
\noindent\textbf{Curating a corpus of research papers.} To curate our corpus for the analysis, we followed five steps.

First, we gathered an \emph{initial} set of 190,790 papers. To ensure comprehensive coverage, we selected venues in which RAI research is typically published (e.g., CHI, CSCW, FAccT, AIES, NeurIPS, ICML). To begin with, we selected publication venues from six core Computer Science areas on Google Scholar: \emph{Artificial Intelligence (AI)}; \emph{Human-computer interaction (HCI)}; \emph{Natural Language Processing (NLP)}; \emph{Database \& Information Systems}; \emph{Data Mining \& Analytics}; and \emph{Computer Vision}. For each area, we included ten highly-ranked venues based on Scholar's h5 index.\footnote{Google Scholar defines the h5 index as the $h$-index calculated for articles published in the most recent five complete years. Specifically, it represents the highest number $h$ such that $h$ articles from the period 2018-2022 each received at least $h$ citations.} Papers from these venues cover more than 88\% of the total citations in each of those six areas (Figure A.6, Supplementary Information).

Second, we extended this initial set with all the papers (N=950) published in the two main conferences dedicated to RAI: FAccT (started in 2018) and AIES (started in 2017). 

Third, we added to this set papers (N=57,754) that had code publicly available on the site ``Papers With Code''.

Fourth, we had to select the timeframe of our analysis. We determined 2015 to be an appropriate starting point. Considering the emergence of RAI research and the establishment of AIES and FAccT conferences around 2017-2018, we defined a timeframe for our analysis from 2015 to 2022. This decision was based on three factors: \emph{(a)} the AI community began witnessing the first significant real-world impacts of deep-learning applications around 2015~\cite{lecun2015deep, silver2016alphago}, and the first incidents of irresponsible AI~\cite{forbes_google, microsoft_bot}; \emph{(b)} establishing a conference usually requires several years of prior planning and discussions through workshops, symposiums, and other platforms. Therefore, 2015 was a conservative choice; and \emph{(c)} the entry of new technology companies such as OpenAI and Hugging Face, which possibly changed the patent landscape among technology conglomerates (e.g., IBM has shifted its focus from patent leadership in 2020~\cite{ibm_patents}). This filtering left us with 177,881 papers.

Finally, we excluded benchmarks, tutorials, surveys, and other non-research publications that are unlikely to end up in patents or repositories. This left us with 174,705 papers.

\smallskip
\noindent\textbf{Retaining RAI papers.} To ensure a systematic way of retaining RAI papers in our corpus, we relied on the NIST AI Risk Management Framework~\cite{nist_framework}. The framework identifies characteristics that contribute to trustworthiness in AI systems. These characteristics are defined as a set of 31 keywords, which describe RAI topics such as fairness, explainability, accountability, privacy, and sustainability. We chose NIST because it is a recent framework from a renowned organization for developing frameworks and standards. Alternatives include the Principled Artificial Intelligence from the Berkman Klein Center~\cite{fjeld2020principled}, which aligns with the NIST framework.

Starting off with the NIST keywords, we found that some of these keywords are non-discriminative. For example, the keywords `validation', `accuracy', and `robustness' are often used as a generic term in AI but not as a keyword that could be used to identify RAI papers (full list of removed keywords is provided in the Supplementary Information). After removing them, we retained 12 keywords. We then added 7 more keywords related to `sustainability' as it is an important topic in the public RAI debate~\cite{strubell2019energy},\footnote{With the increasing power consumption needed to train AI systems, sustainability is becoming a critical focus area. This involves considering the environmental impact of AI systems, such as the energy consumption of data centers and the carbon footprint of training large models~\cite{strubell2019energy}. The need for sustainable AI practices extends beyond environmental concerns to encompass long-term viability and responsible resource usage in AI development and deployment. and 8 more keywords related to governance and accountability---topics covered by regulations (e.g., EU AI Act).} We then queried with the 27 keywords. During this process, we noticed spurious matches, prompting us to remove 2 keywords leading to these irrelevant matches. This process led to a final set of 25 keywords 
(Table A.4, Supplementary Information).

Using ST5~\cite{ni2021sentence}, a deep learning natural language processing framework, we generated embeddings for both the chosen NIST keywords and the abstracts of the papers, and then calculated the pairwise similarity between them. For each paper, we assigned the most similar keyword using the highest cosine similarity between the two embeddings, resulting in a one-to-one pairing of papers and topics. To ensure a precise yet conservative approach, we selectively included papers with a similarity score above the 99\textsuperscript{th} percentile---a less relaxed threshold (e.g., 95\textsuperscript{th}) could also have been used~\cite{maclaughlin2021recovering}, but a stricter one was preferred to ensure no spurious matches. However, this threshold presented a trade-off between precision and recall. In our case, where the focus is on identifying the most impactful RAI contributions, precision was prioritized over recall to avoid including a broader range of impactful papers from AI rather than specifically from RAI.

Finally, we eliminated duplicate papers and retained high-quality papers by filtering out those with a number of citations lower than the number of years since publication.

\smallskip
\noindent\textbf{Retrieving citations in patents.}
We retrieved citations in patents in three steps. First, we connected Semantic Scholar's paper metadata with USPTO patents. Second, we used the Grobid\footnote{https://github.com/kermitt2/grobid} deep-learning framework to extract titles and authors from patent references systematically. We then generated the embeddings of the extracted titles by using the ST5 model~\cite{ni2021sentence}.
Once we obtained the embeddings, we computed the pairwise cosine similarity to identify potential title matches, setting a distance threshold at 0.06. This threshold was chosen using the elbow rule, which involves plotting the number of title matches as a function of the distance threshold and selecting the elbow of the curve as the optimal threshold.
Third, to ensure precision, we cross-referenced author names from Semantic Scholar against those in patent references using the Levenshtein distance metric~\cite{marx2022reliance}, setting a similarity threshold above 0.8 through an elbow strategy. Our approach effectively overcomes the limitations found in other methods, such as a rule-based scoring technique~\cite{marx2022reliance} that suffers from reproducibility and adaptability issues to new datasets, or a purely machine learning-based method (Biblio Glutton) that is not scalable due to extensive pairwise comparisons.

\smallskip
\noindent\textbf{Final Datasets Statistics.}
The final list of RAI papers consists of 1,747 papers (Table~\ref{tab:rq1}), of which 557 (31.9\%) are about fairness, 538 (30.8\%) cover privacy, 318 (18.2\%) discuss accountability, 219 (12.5\%) focus on explainability, and 115 (6.6\%) deal with sustainability.

\subsection{Limitations: Data, Metrics, and Analysis}
\label{subsec:limitations}
We acknowledge six limitations. First, the total number of papers analyzed may not represent the full extent of published papers as we relied on Semantic Scholar. Nevertheless, Semantic Scholar has been used in similar analyses~\cite{cao2023breaking} with coverage comparable to Google Scholar~\cite{hannousse2021searching}. To mitigate potential biases, future research could replicate our method using alternative databases such as Crossref\footnote{https://www.crossref.org} and OpenAlex.\footnote{https://openalex.org} Second, our title-matching method for associating Semantic Scholar papers with patent references may not be exhaustive. Future studies could explore alternative matching algorithms such as Biblio Glutton\footnote{https://github.com/kermitt2/biblio-glutton} and \citet{marx2022reliance}'s.
Third, our approach does not capture the intent for a citation. When we attempted to identify specific intents by searching for mentions of particular unigrams such as ``extension'', ``future'', and ``use'' in citations (similar to what was done in previous work~\cite{cohan}), we encountered two main challenges: 1) establishing a ground truth was difficult and relied on our estimations of potential intent; and 2) even with a preliminary ground truth, the results were not promising (F1 = 0.67 for intent classification as reported in~\cite{cohan}). To this end, we deliberately decided not to capture the intent for citations. Fourth, as we shall see next, our metrics for assessing research impact are limited to data from USPTO (patents) and GitHub (repositories). Future studies could expand to a wider range of sources (e.g., patents from the European Patent Office), enhancing the comprehensiveness of impact assessment. Fifth, as RAI is an emerging area, we acknowledge that the number of papers may currently be limited; however, this can change drastically in the future. Finally, as we shall see next, our metrics have limitations, particularly in capturing broader societal impact. Future work could investigate the overlap between research studies that shape policy and those that influence patents. For example, one may examine how research translates into legislation by comparing RAI paper topics with legislative drafting such as the EU AI Act or the NIST's AI RMF, or by examining whether research adequately covers real-world incidents by comparing RAI paper topics with databases collating AI incidents~\cite{ai_incidents_db}. Future research could also replicate our methodology to study how authors' diversity (e.g., nationalities, backgrounds, and areas of expertise) translates to varying levels of impact~\cite{septiandri2023weird, septiandri2024western}.

\subsection{Metrics}
\label{subsec:metrics}
We chose our metrics because they align with peer-reviewed metrics~\cite{uzzi2013atypical, ahmadpoor2017dual, manjunath2021comprehensive, cao2023breaking}. These metrics are applied either to \emph{patents} or \emph{repositories}. To simplify notation, we use the variable $X$ to indicate one of the two sets of patents or repositories (Table \ref{tab:definitions}). Note that we analyzed patents and repositories separately because if we were to combine them, we would only study papers that go into repositories and cancel any patent effects in our dataset. Additionally, if we were to combine them, we would have left with only 63 papers that both go into papers and repositories; a number that is insufficient for any quantitative analysis that divides these papers across the five RAI topics. 
\smallskip

\begin{table}[t!]
\centering
\caption{We used these variables throughout our analysis to measure the impact of  RAI research on innovation (through patents) and development (through code repositories).}
\label{tab:definitions}
\scalebox{.94}{
\setlength{\tabcolsep}{1.25pt}

\begin{tabular}{l|l}
\toprule
\textbf{Variable} & \textbf{Description} \\ 
\midrule
$P$ & Patents in the corpus (1980-2022, USPTO). \\
$R$ & Repositories in the corpus (2008-2022, GitHub). \\
$X$ & Generic set of citing entities; could be $P$ or $R$. \\
$U$ & Papers in the corpus (2015-2022). \\
$H$ & Papers from the venues under study. \\
$H'$ & Papers not from the venues under study. \\
$c_{\mathrm{in}}(y)$ & Papers associated with a document $y$ (paper/patent). \\
$c_{\mathrm{out}}(u)$ & Documents (among papers/patents) citing paper $u$. \\
$H \mid X$  & Papers in $H$ that went into patents/repositories. \\
$(H \mid X)'$  & Papers in $H$ that didn't go into patents/repositories. \\
$H' \mid X$  & Papers in $H'$ that went into patents/repositories. \\
$X \mid H$  & Patents or repositories that cite papers in $H$. \\
$U \mid H$  & Set of papers that cite papers in $H$. \\
$r(H) $ & Proportion of papers in $H$ that went into patents. \\
$r(H')$ & Proportion of papers in $H'$ that went into patents. \\
\bottomrule
\end{tabular}
}

\end{table}



\noindent\textbf{Measuring the impact of RAI papers on patents and repositories.}
To quantify RAI papers' research impact, we used patent citations\footnote{Because of the small number of papers that go into patents, our analysis lacks sufficient statistical power to differentiate impacts based on the number of patent citations. Therefore, we focused on papers that have received at least one patent citation.} and the creation of GitHub repositories.\footnote{Based on data from `Papers with Code', the median number of repositories linked to each paper is 1. Consequently, our analysis focuses on the impact of papers that are associated with at least one repository, without distinguishing between papers linked to multiple repositories.} A Z-test was conducted to evaluate the difference between $r(H \mid X)$ and $r(H' \mid X)$, defined as follows: $r(H) = \frac{\vert (H \mid X) \vert}{\vert H \vert}$, and $r(H') = \frac{\vert (H' \mid X) \vert}{\vert H' \vert}$, where $H$ is the set of papers from the venues under study; $H | X$ is the set of papers in $H$ that went into patents (or repositories) $X$; $H'$ is the complement set of papers (i.e., all papers in the venues other than those considered in $H$); and $H' | X$ is the set of papers in $H'$ that went into patents (or repositories). A straightforward extension to count the number of total citations (denoted as $c(\cdot)$) in the $H$ set is $r(H) = \frac{c(H \mid X)}{c( H )}$.

\sloppy To compare academic impact, we conducted an unpaired $t$-test comparing the mean citation counts of papers involved in patents (or repositories) $\mu_{H \mid X}$, with those that were not $\mu_{(H \mid X)'}$. Here, $\mu_{H \mid P}$ is calculated as $\frac{1}{|(H \mid X)|}\sum_{h\in (H \mid X)}{c_{\mathrm{out}}(h)}$, and $\mu_{(H \mid X)'}$ is $\frac{1}{|(H \mid X)'|}\sum_{h in (H \mid X)'}{c_{\mathrm{out}}(h)}$.
\smallskip

\noindent\textbf{Measuring the time it takes for RAI papers to have an impact on patents and repositories.} To measure the time that it takes for RAI papers to have an impact on patents and repositories, we used the survival analysis model Kaplan-Meier estimator~\cite{dudley2016introduction}. We selected this model because the analysis presents a time-to-event challenge with papers that are \emph{right-censored} (e.g., the first association with a patent might not happen when we intend to make conclusions).

The survival function $S(t)$, which calculates the likelihood that a paper will remain unassociated with a patent beyond a given time $t$, is estimated non-parametrically by $\hat{S}(t) = \prod_{i: t_i \le t} \left( 1 - \frac{d_i}{n_i} \right)$. Here, $t_i$ is a point in time when an association with a patent occurs, $d_i$ counts the occurrences of these first associations at $t_i$, and $n_i$ is the number of papers that have neither been associated with a patent nor censored by time $t_i$.
\smallskip

\noindent\textbf{Identifying the top institutions that produce RAI papers that make an impact on patents and repositories.} We took all the papers that went into patents or repositories, associated these papers with their co-authors' distinct institutions, and summed the total number of resulting papers at each institution. This allowed us to pinpoint the leading institutions with the highest impact. We considered the top 50 institutions by total number of papers and then ranked them based on the average percentage of papers that went into patents or repositories across all research areas. Note that we counted each paper for every institution its authors are affiliated with. Although fractional allocation was considered as an alternative, we chose the whole number assignments for ease of interpretation, particularly because this method did not change the rankings. These numbers can be regarded as the upper bound of the impact.
\smallskip

\noindent\textbf{Identifying whether combining conventional and unconventional knowledge in RAI papers affects their impact.}
Using the methodology proposed by \citet{uzzi2013atypical} in a Science publication, we studied how conventional and unconventional knowledge affects the impact of RAI papers on patents and repositories.
We measured the conventionality of a paper by analyzing the pairwise combinations of references in its bibliography, factoring in the frequency of citations. Therefore, the conventionality score for each paper was calculated from the possible combinations of cited sources. For each pair of venues in the references, we computed both the actual co-occurrence in our dataset and the expected co-occurrence in a hypothetical scenario, using a method akin to the randomization strategy outlined by \citet{uzzi2013atypical}. This method preserved two parameters: the original distribution of citations and the years of publication. To align with these parameters, we tagged each source pair with the publication year of the referencing paper, represented as: ${ (\mathrm{venue}_i, \mathrm{venue}_j, \mathrm{year}) \cdots }$. In our randomization process, we permuted the entries for $\mathrm{venue}_j$, matching them with entries from corresponding years. By executing this permutation process 1,000 times, we created a null model that accurately mirrored the structure of the citation network, ensuring a consistent citation frequency across papers.

We standardized the conventionality scores to center the values at zero (i.e., $z$-scores). Consequently, scores under 0 indicate less conventional references, whereas scores above 0 indicate more conventional ones. Each paper is linked with a set of $z$-scores, each corresponding to a pair of cited venues. Essentially, when a RAI paper references well-known conferences like NeurIPS, ICML, and ICLR, it is considered more conventional (scoring above 0). In contrast, citing less common conferences like The Web Conference, CIKM, and CHI is considered unconventional, leading to scores below 0. Conservatively, in line with \citet{uzzi2013atypical}'s methodology, we selected the 10\textsuperscript{th} percentile from the set of $z$-scores for each paper. This score is the lowest conventionality score that only 10\% of the papers achieve.

\section{Results}
\label{sec:results}
\subsection*{\textbf{Q}\textsubscript{1}: Which Types of RAI Papers Go Into Patents and Repositories?}

\begin{figure}[t!]
    \centering
    \includegraphics[width=1\columnwidth]{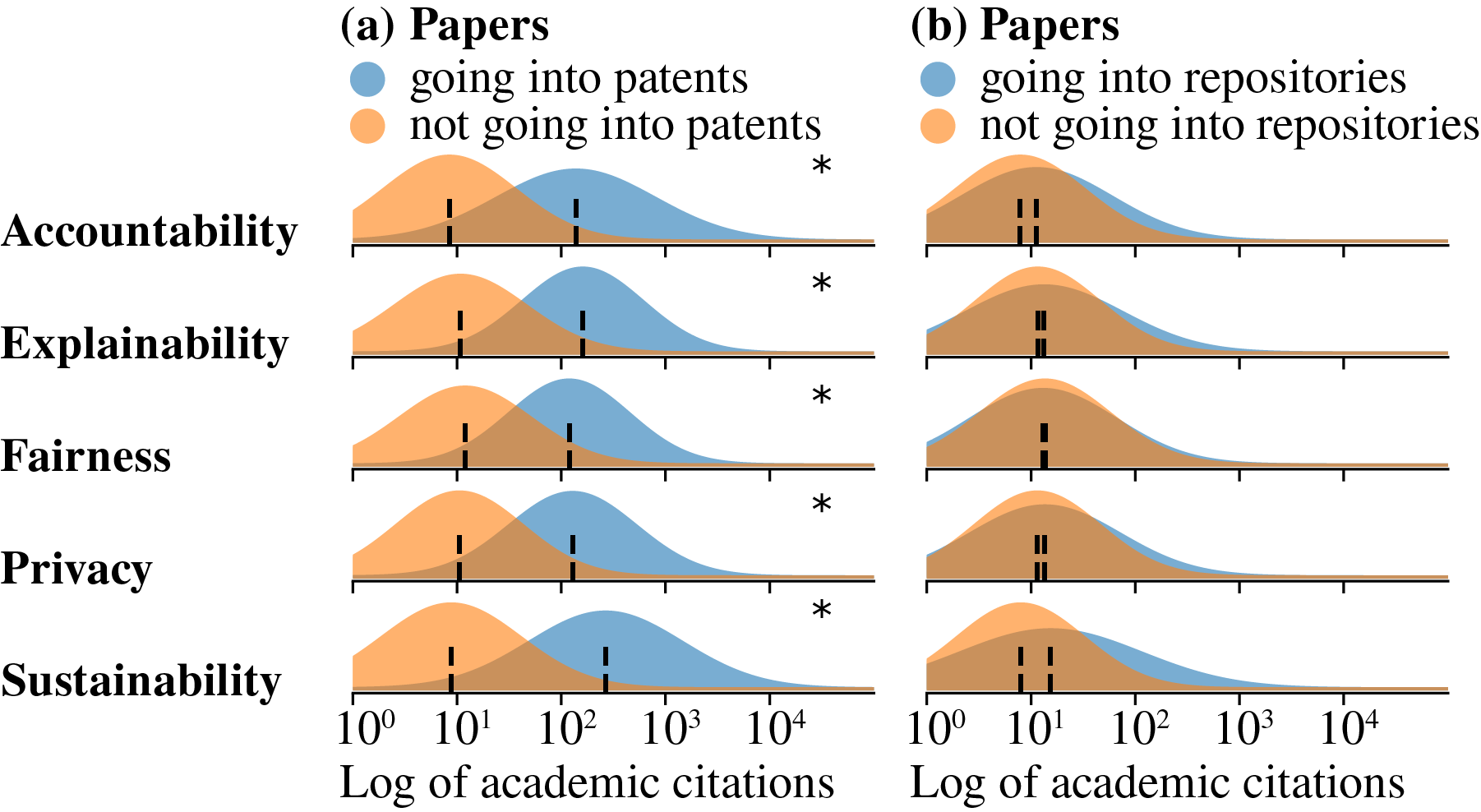}
    \caption{Difference in academic citations of RAI papers going into patents and repositories. Each row represents a RAI topic, with the number of academic citations shown on a logarithmic scale. \emph{(a)} Distributions of papers going into patents (blue) vs. those that do not (orange); and \emph{(b)} distributions of papers going into repositories (blue) vs. those that do not (orange). Wider gap between the distributions in each plot indicates that papers going into patents tend to attract more academic citations that those that do not. Statistically significance difference between the means of the two distributions is at 0.001 level, and marked with a $*$.}
    \label{fig:academic-citations}
\end{figure}

\noindent{\textbf{A small number of highly-cited RAI papers went into patents and repositories.}}
Despite a mere 7.2\% of Sustainability (out of 111) papers going into patents (Table~\ref{tab:rq1}), these papers disproportionately received 52.2\% of academic citations. Half of the papers going into patents in Sustainability are seminal papers with more than 400 citations each, such as ``\emph{Hidden Technical Debt in Machine Learning Systems}''~\cite{sculley2015hidden} and ``\emph{Energy and Policy Considerations for Deep Learning in NLP}''~\cite{strubell2019energy}. Similarly, there are only 6.9\% of Privacy (out of 533) papers going into patents, but they received as much as 44.5\% of academic citations (Figure~\ref{fig:academic-citations}a).  On the other hand, there are 52.3\% of Sustainability and 52.9\% of Privacy papers going into repositories, and they received 82.5\% and 72.3\% of academic citations. Interestingly, the impact is lower in Fairness papers, where there are 4.9\% papers going into patents, but they received 25.5\% of academic citations (Figure~\ref{fig:academic-citations}b).

To provide context for these findings, we compared the percentage of RAI papers that went into patents or code repositories with similar metrics from other AI subfields such as Computer Vision and NLP. By computing our metrics on all papers in our dataset (whether RAI or not), we found that approximately 16.9\% of papers in Computer Vision and 18.7\% in NLP went into patents or code repositories and received 30.6\% and 32.4\% of academic citations, respectively. The lower rates for RAI papers compared to other AI subfields can be attributed either to the relatively young yet exponentially growing nature of RAI research, or to the possibility that more translational work is needed.

\begin{table*}[t!]
    \centering
    \caption{The impact of RAI papers on patents and repositories. Each row represents a RAI topic, with the numbers in parentheses calculated from the values outside the parentheses. These numbers indicate: (1) the percentage of papers going into patents, (2) the percentage of papers going into repositories, and the percentage of academic citations for (3) papers going into patents and (4) papers going into repositories, relative to the total academic citations for each topic.}
    \label{tab:rq1}
    \scalebox{0.94}{
        \begin{tabular}{c|ccc|ccc}
        \toprule
        \textbf{Topic} & \shortstack[c]{\textbf{Papers going into} \\ \textbf{patents}} & \shortstack[c]{\textbf{Papers going into} \\ \textbf{repositories}} & \textbf{Papers} & \shortstack[c]{\textbf{Number of} \\ \textbf{academic citations} \\ \textbf{of papers going into} \\ \textbf{patents}} & \shortstack[c]{\textbf{Number of} \\ \textbf{academic citations} \\ \textbf{of papers going into} \\ \textbf{repositories}} & \shortstack[c]{\textbf{Number of} \\ \textbf{academic}  \\ \textbf{citations}} \\
        \midrule
        Fairness & 27 (4.8\%) & 264 (47.4\%) & 557 & 6066 (25.5\%) & 13746 (57.7\%) & 23829 \\
        Privacy & 40 (7.4\%) & 283 (52.6\%) & 538 & 12658 (47.6\%) & 19232 (72.3\%) & 26605 \\
        Accountability & 9 (2.8\%) & 132 (41.5\%) & 318 & 4451 (37.9\%) & 7887 (67.2\%) & 11739 \\
        Explainability & 12 (5.5\%) & 121 (55.3\%) & 219 & 3900 (26.2\%) & 11843 (79.7\%) & 14861 \\
        Sustainability & 9 (7.8\%) & 60 (52.2\%) & 115 & 4859 (52.0\%) & 7788 (83.3\%) & 9348 \\
        \bottomrule
        \end{tabular}
    }
\end{table*}

\subsection*{\textbf{Q\textsubscript{2}}: How Long Does It Take for RAI Papers To Go Into Patents or Repositories?}
\begin{figure}[t!]
\centering
\includegraphics[width=1\columnwidth]{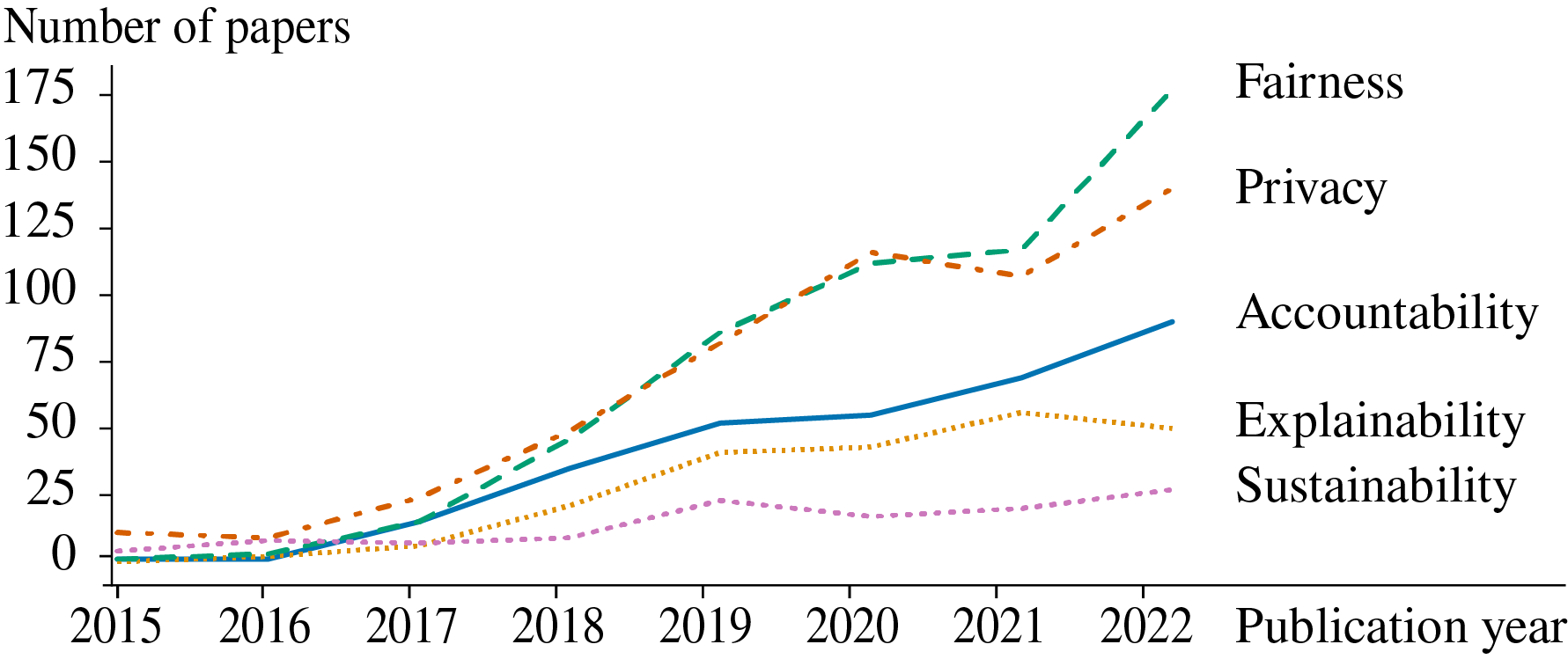}
\caption{Number of papers for each RAI topic between 2015-2022. Since 2017, there has been an increasing trend of RAI papers, especially in Fairness and Privacy.}
\label{fig:yearly}
\end{figure}

\noindent{\textbf{Increasing popularity of papers in Fairness and Privacy.}} In all topics, there has been an upward trend of RAI papers (Figure~\ref{fig:yearly}). While Privacy has been at the top since 2015, Fairness has followed an upward trend since 2017. Sustainability, Explainability, and Accountability remained constant over the years.

\begin{figure}[t!]
\centering
\includegraphics[width=1\columnwidth]{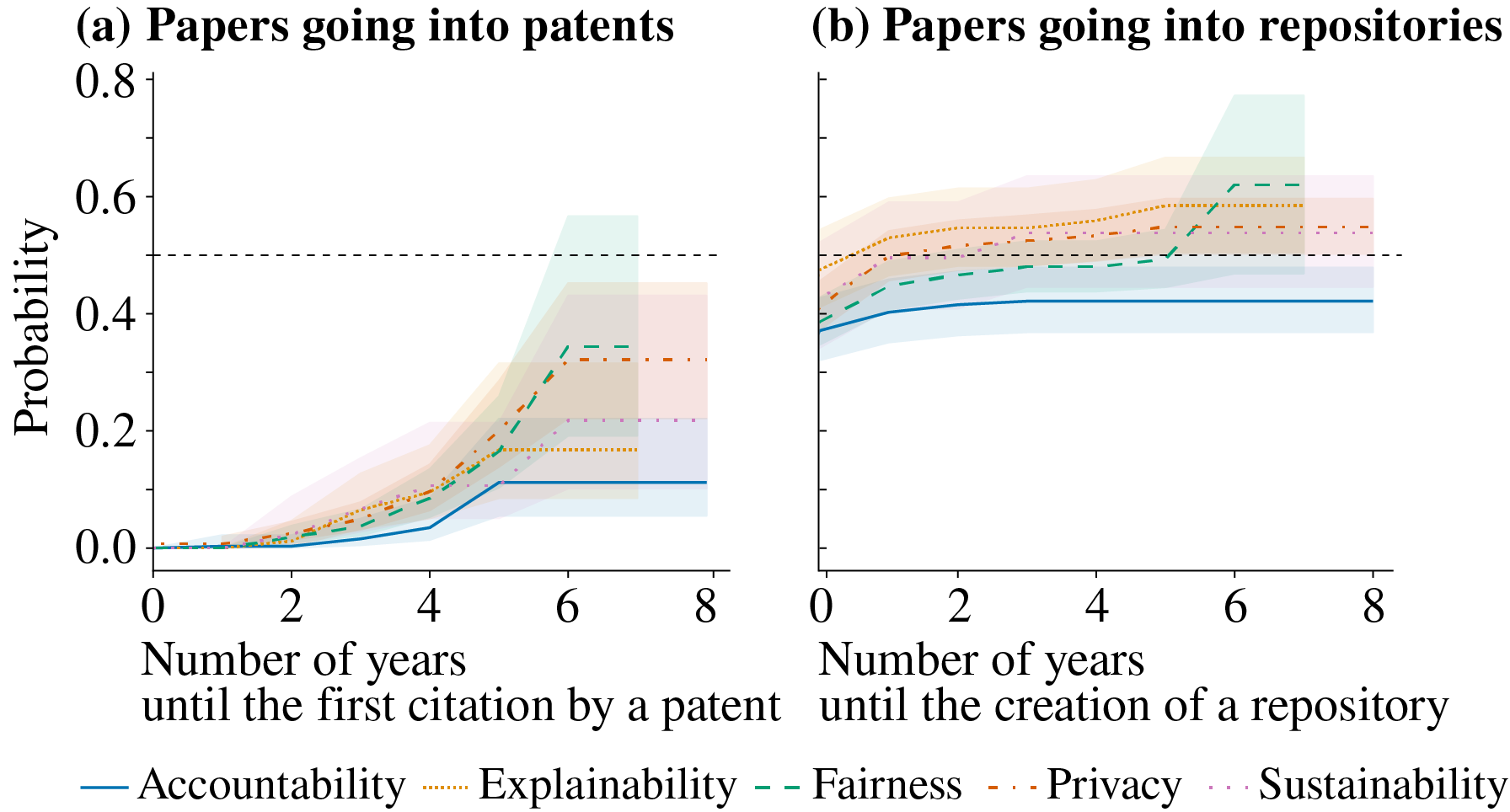}
\caption{Survival analysis of papers going into patents and repositories. For each topic, the probability of its papers going into: (a) patents for the first time (i.e., receiving the first citation), and (b) repositories for the first time (i.e., receiving the first star or fork on GitHub). The horizontal line at 0.5 represents the threshold beyond which papers have a better chance of going into patents or repositories than a coin flip.}
\label{fig:survival}
\end{figure}

\smallskip
\noindent{\textbf{The time lag between a paper going into a repository is far shorter than that of a paper going into a patent.}} On average, it takes more than 8 years for RAI papers to go into patents. In contrast, it only takes 1 year for them to go into repositories, though it can take more than 5 years for Fairness and Accountability papers (Figure~\ref{fig:survival}).

Two potential reasons might explain this lag. The first is related to the longer process of publishing patents compared to papers. This leads to a situation where it might take up to a decade for a paper to accumulate patent citations~\cite{cao2023breaking}. The frequency with which a paper is cited in a patent is influenced by the time elapsed since its publication. To address time dependency, we conducted a survival analysis, as detailed in \S\nameref{subsec:metrics}. As observed in Figure~\ref{fig:survival}a, the probability of a paper going into patents increases with each passing year after its publication.
However, it might also be due to the overproduction of papers on certain topics (e.g., Fairness or Privacy), which exceeds the rate of patent production. In simpler terms, if a topic becomes overly popular, it might lead to an oversaturation of papers, thus lowering the probability of a paper going into patents. This can create a bottleneck effect, straining the already lengthy patenting process, which typically takes 24-30 months from filing to grant~\cite{uspto_research}.

\subsection*{\textbf{Q\textsubscript{3}}: Which Institutions Produce RAI Papers That Go Into Patents or Repositories?}

\begin{table*}[t!]
    \centering
    
    \caption{Top 20 institutions ranked by the total number of RAI papers going into patents and repositories. The `Topic diversity' column shows the Gini-Simpson index for the total of RAI papers across the five RAI topics. The higher the topic's diversity value, the greater the diversity among topics. The absence of papers going into patents at two institutions (e.g., Tsinghua University and University of Virginia) highlights the need to consider alternative measures of impact (e.g. repositories).} 
    \label{tab:institutions}
    \scalebox{0.94}{
    \begin{tabular}{l|ccc|cc}
    \toprule
    \textbf{Institution} & \shortstack[c]{\textbf{Papers going into} \\ \textbf{patents}} & \shortstack[c]{\textbf{Papers going into} \\ \textbf{repositories}} & \textbf{Total} & \shortstack[c]{\textbf{Topic} \\ \textbf{diversity}} & \shortstack[c]{\textbf{Total} \\ \textbf{RAI papers}} \\
    \midrule
    National University of Singapore & 1 & 16 & 17 & 0.66 & 28 \\
    Google (United States) & 4 & 11 & 15 & 0.71 & 31 \\
    Cornell University & 5 & 9 & 14 & 0.71 & 22 \\
    Carnegie Mellon University & 2 & 11 & 13 & 0.74 & 32 \\
    Stanford University & 1 & 8 & 9 & 0.70 & 21 \\
    Imperial College London & 2 & 7 & 9 & 0.69 & 8 \\
    Massachusetts Institute of Technology & 3 & 5 & 8 & 0.75 & 13 \\
    Hong Kong University of Science and Technology & 2 & 6 & 8 & 0.72 & 13 \\
    University of Oxford & 3 & 5 & 8 & 0.69 & 11 \\
    Nanyang Technological University & 1 & 6 & 7 & 0.63 & 17 \\
    Microsoft (United States) & 2 & 5 & 7 & 0.66 & 16 \\
    University of California, Berkeley & 1 & 6 & 7 & 0.68 & 14 \\
    University of Wisconsin–Madison & 1 & 6 & 7 & 0.68 & 10 \\
    ETH Zurich & 1 & 5 & 6 & 0.71 & 16 \\
    Tsinghua University & 0 & 6 & 6 & 0.53 & 11 \\
    Helmholtz Center for Information Security & 1 & 5 & 6 & 0.00 & 6 \\
    University of Virginia & 0 & 5 & 5 & 0.58 & 10 \\
    University College London & 2 & 3 & 5 & 0.53 & 8 \\
    University of California, Los Angeles & 1 & 4 & 5 & 0.72 & 8 \\
    Zhejiang University & 1 & 4 & 5 & 0.41 & 7 \\
    \bottomrule
    \end{tabular}
    }
    
\end{table*}

\noindent\textbf{Top institutions with papers going into patents and repositories are Cornell University and National University of Singapore (NUS), respectively.} Carnegie Mellon University (CMU) produces the most RAI papers across all topics (Table~\ref{tab:institutions}). However, NUS is particularly prominent when considering papers going into repositories, with significant contributions in Fairness and Privacy. Overall, universities tend to have a greater impact on RAI than companies. However, when grouping the number of RAI papers by topic, we observed that companies were also leading in certain areas. In Accountability, Amazon and Google were among the top institutions, along with Stanford University and Oxford University; in Explainability, Microsoft shares the top places with TU Berlin; and in Fairness, Google leads along with NUS and CMU.

Interestingly, the impact of top U.S. institutions is not exclusive. Among the top 20 institutions, 10 are in the U.S., 5 are in Europe, and 5 are in Asia. This paints a clear picture that a significant number of papers going into patents also come from European and Asian institutions. This indicates a more global distribution of influence than previously recognized~\cite{hbr_china_emerging}.

When looking at the topic diversity (measured as the Gini-Simpson index~\cite{jost2006entropy}), the top institution is Massachusetts Institute of Technology, followed closely by Carnegie Mellon University, University of California, Los Angeles, Hong Kong University of Science and Technology, and Google (United States), all with values greater than 0.7; a value that ranges from 0 to 1, where 1 indicates greatest diversity. We also checked whether there is any association between topic diversity and the number of papers that go into patents or repositories. We found that topic diversity is weakly correlated with the number of papers going into repositories (Pearson's $r = 0.24, p < 0.05$), while the correlation with the number of papers going into patents was not statistically significant. 

\subsection*{\textbf{Q\textsubscript{4}}: How Does Combining Conventional and Unconventional Knowledge in RAI Papers Affect Their Impact?}

\begin{figure}[t]
\centering
\includegraphics[width=1\columnwidth]{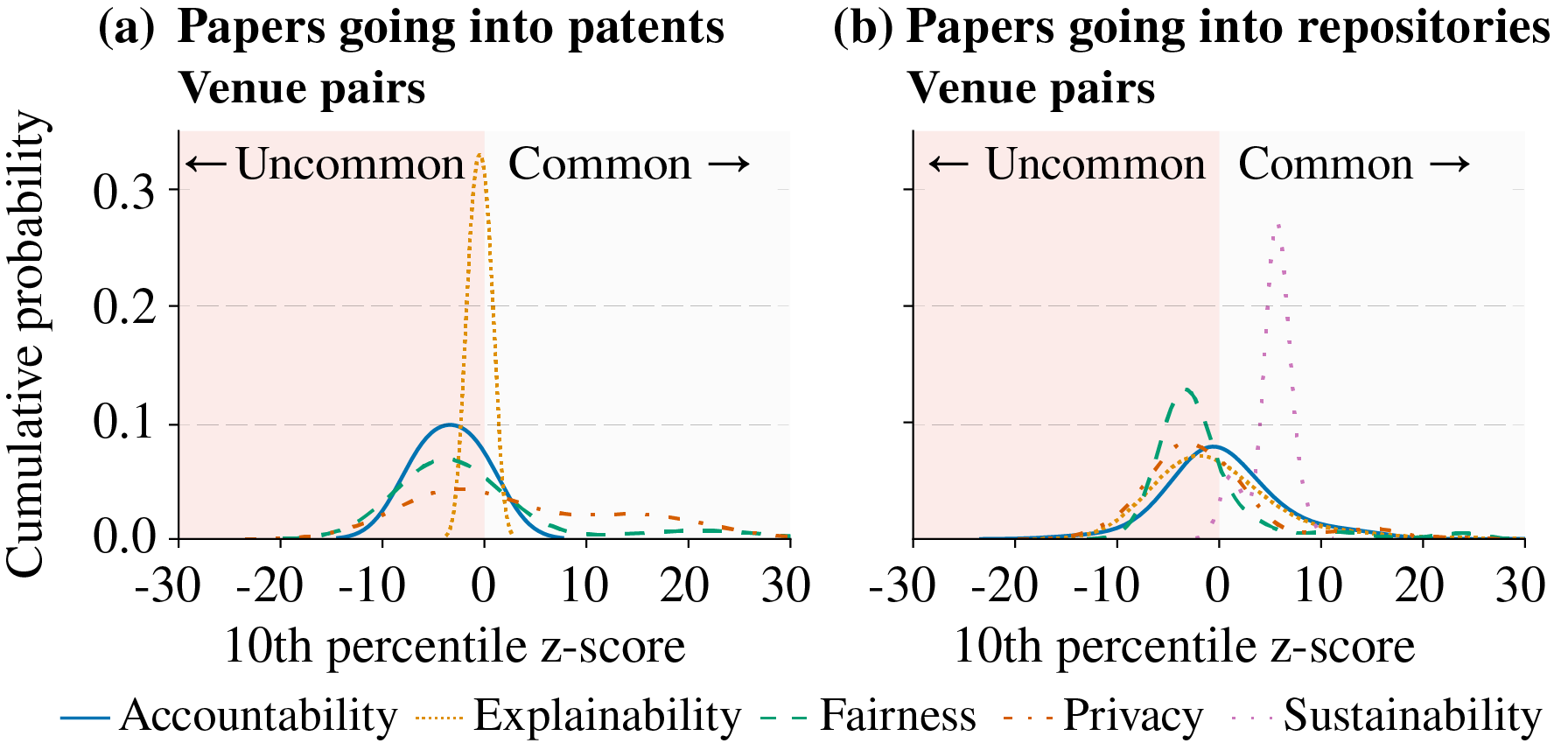}
\caption{Conventionaliy of RAI papers going into patents and repositories. A topic's conventionality is based on whether its papers---those that going into patents (left plot) and those going into repositories (right plot)---have (un)usual combinations of citations. A larger red area indicates a higher degree of unconventionality. Papers going into patents in Accountability often cite papers discussing AI applications across different venues (e.g., KDD, ICLR, USENIX, CVPRW, NeurIPS).}
\label{fig:conventionality}
\end{figure}

The conventionality score of a paper is determined by examining every possible pair of its references. This score indicates whether its references are typically or rarely seen in other papers. The frequency of each reference pair is noted as it appears in the dataset and the frequency expected under random conditions (null model described in \S\nameref{subsec:metrics}).

\smallskip
\noindent\textbf{RAI papers going into patents or repositories have unconventional combinations of prior work.} In general, papers going into patents or repositories are grounded on unconventional combinations of prior work. However, Privacy papers that went into patents can be more conventional (Figure~\ref{fig:conventionality}a), similar to Sustainability papers that went into repositories (Figure~\ref{fig:conventionality}b). To see this distinction in a concrete example, consider two unconventional papers: \emph{``Alleviating Privacy Attacks via Causal Learning''} by~\citet{tople2020alleviating} and \emph{``Practical One-Shot Federated Learning for Cross-Silo Setting'' by~\citet{li2021crosssilo}}. These papers draw from uncommon pairs of venues such as NeurIPS and IEEE Computer Security Foundations Symposium, and Conference on Computer and Communications Security and ICLR and have consequently gone into several repositories. While security and machine learning papers might be more similar in their approach, let us consider another example of two unusual papers in Accountability: \emph{``Data-Centric Explanations: Explaining Training Data of Machine Learning Systems to Promote Transparency''} by~\citet{anik2021data} and \emph{``Amortized Generation of Sequential Algorithmic Recourses for Black-Box Models''} by~\cite{verma2022amortized}. These papers draw from uncommon pairs of venues such as HCI and ICML. On the other hand, we found that Sustainability papers going into repositories are more likely to be grounded on conventional combinations of prior work.
\section{Discussion}
\label{sec:discussion}

\subsection{Main Findings}
By analyzing 1,747 papers in the field of RAI from leading Computer Science conferences, we found that a small subset that goes into patents or repositories is highly cited. The majority of these papers that go into patents focus on sustainability and privacy, while those that go into repositories focus on explainability. The temporal dynamics between papers that go into patents and those that go into repositories vary. The time frame for papers that go into patents is 8 years, while those that go into repositories are 1 year. Our survival analysis revealed not only the presence of a rapid publication cycle and a slower patent production cycle, but also that these cycles vary across different RAI topics. This suggests the need to reevaluate existing assumptions about the pace and consistency of knowledge production in both patents and repositories. When examining the top institutions producing impactful RAI papers in terms of patents and repositories, we observed that the impact of top U.S. institutions is not exclusive. Although our study concentrated on patents in the U.S., we noticed a significant contribution from European and Asian institutions. This indicates a more global distribution of influence than previously recognized. Finally, RAI papers generally rely on unconventional knowledge, often integrating expertise from fields such as ML and HCI, in both patents and repositories.

Our metrics reveal that the impact of patents and repositories extends beyond the authors themselves in four ways. First, it signals authors' intentions. Authors often file patents or create repositories to demonstrate real-world impact. Second, it surfaces the trade-offs between costs and patentability. Filing a patent incurs costs, but is often driven by commercialization and intellectual property protection. While not all papers are patentable, those that are may still represent significant technological advancements. Third, it illustrates the shift towards open science. Many companies value open-source repositories for fostering collaboration and innovation, even if it means disclosing some aspects of their technology. Fourth, it shows the trade-offs between economic benefits and knowledge democratization. Patents allow companies to commercialize inventions by granting them temporary monopolies, while repositories democratize knowledge by offering free access to information.

\subsection{Implications}
By synthesizing our findings with those documented in prior literature, we propose three recommendations for enhancing the translational landscape of RAI research.
\smallskip

\noindent\textbf{Developing Initiatives Between Academia and Industry.} Translational research is crucial in bridging the gap between scientific breakthroughs and their practical application~\cite{searles2016approach}. Echoing the findings of previous studies~\cite{cao2023breaking}, our study suggests that only a portion of RAI research significantly contributes to patents and repositories.
Therefore, it is beneficial to foster collaborative efforts between academia and industry. This can be achieved through various initiatives. First, by establishing dual mentoring programs that connect RAI researchers with mentors from both academic and industrial sectors, including patent professionals. Second, by organizing joint conferences and patent clinics that bring together AI researchers, patent engineers, and industry experts, researchers can gain immediate access to professional guidance on patenting. Finally, by promoting corporate participation in academic endeavors through hackathons and sponsorships~\cite{briscoe2014digital}, with the goal of enhancing the translation of RAI research into practical innovations.
\smallskip

\noindent\textbf{Balancing and Diversifying Focus in RAI Research.} The observed delay between RAI papers and patents may suggest an overemphasis on certain research topics. While this focus on specific areas (e.g., Fairness and Accountability) advances theoretical knowledge, it might inadvertently slow down the development of practical innovations~\cite{mckinsey_ai}. However, slowing down is not necessarily bad. In fact, incremental approaches could be favorable when considering the impact of technical work on social change~\cite{abebe2020roles}. This is advised because otherwise researchers may fail to recognize how changes in behaviors of pre-existing social systems can be caused by the introduction of technology. It can also lead to technical solutions focused purely on ethical or value-aligned deployments, without considering whether a given system functions and provides benefits~\cite{raji2022fallacy}, or even a technical solution is needed at all~\cite{selbst2019fairness}. To address these challenges, a more strategic approach is necessary to accelerate the transition from academic discoveries to patentable technologies. This would involve reevaluating RAI research priorities to ensure they align with practical needs. Our findings also suggest that papers that go into patents tend to concentrate on sustainability and privacy, whereas those in repositories are focused on explainability. This discrepancy highlights the importance for academia to realign its research priorities with industry demands.
\smallskip

\noindent\textbf{Fostering Cross-discipline RAI Research.} The impact of RAI papers is shaped by how they build upon prior knowledge. Generally, RAI papers tend to use unconventional combinations of prior knowledge, and often draw upon expertise from diverse fields such as ML and HCI. To continue making a significant impact on both innovation and development, it is important to encourage researchers to engage in cross-disciplinary research efforts. This can be achieved by the formation of research teams that include experts from multiple disciplines~\cite{rakova2021responsible} (e.g., computer science, social sciences, and law), incorporating diverse perspectives and technical expertise~\cite{moitra2022ai} as well as contributing towards organizational culture shift~\cite{rakova2021responsible}.  Additionally, educational programs that promote interdisciplinary work~\cite{frodeman2011interdisciplinary} are required to train AI researchers in ethical, legal, and social implications of their work~\cite{mcdonald2020intersectional}.

\section{Conclusion}
\label{sec:conclusion}
By analyzing 1,747 RAI papers, we found that RAI research, despite being an emerging field, contributes to patents and development. This translation process typically occurs within a time frame of 1 to 8 years, highlighting a gap between research output and translational outcomes. Our findings also challenge the conventional notion that impactful RAI research is predominantly the domain of top U.S. institutions. Instead, European and Asian institutions are also major contributors, indicating a more globally distributed expertise in the field. Finally, the multidisciplinary nature of RAI papers, which often build upon unconventional combinations of prior knowledge, is vital for ensuring the development of fair, transparent, and accountable AI systems.

\subsection{Researcher Positionality Statement and Ethical Considerations}
\label{subsec:positionality}
The research team includes three men from Southeast Asia and Southern Europe, working in industry. Our shared backgrounds include Human-Computer Interaction, software engineering, AI, social and ubiquitous computing, and urbanism. In light of recent calls for RAI research to include impact and ethical statements~\cite{olteanu2023responsible}, our research neither involves any subjects nor do we foresee any significant harm that could result.

\appendix
\counterwithin{table}{section}
\counterwithin{figure}{section}

\bibliography{main}

\clearpage

\renewcommand{\thetable}{A.\arabic{table}} 
\setcounter{table}{3} 

\renewcommand{\thefigure}{A.\arabic{figure}} 
\setcounter{figure}{5} 

\section*{Appendix}

\section{Keywords to Filter RAI Papers}
\label{appendix:keywords}

We developed a set of 25 keywords based on the NIST framework to identify RAI papers in our corpus. For all the listed keywords in Table~\ref{tab:nist-keywords}, we prepended ``artificial intelligence'' or ``machine learning'' to ensure the resulting papers are about RAI.

\begin{table}[htbp]
    \centering
    
    \begin{tabular}{l|p{5cm}}
        \toprule
        \textbf{Topic} & \textbf{Keywords} \\
        \midrule
        Fairness & fairness, equality, equity, equitable \\
        Privacy & privacy, anonymity, confidentiality, confidential \\
        Explainability & explainability, explainable \\
        Accountability & accountable, accountability, transparency, auditability, governance, compliance, accountability mechanisms, algorithmic accountability \\
        Sustainability & green, energy-efficient, carbon footprint, environmental impact, eco-friendly, energy consumption, green computing \\
        \bottomrule
    \end{tabular}
    \caption{Selected NIST keywords to filter RAI papers. From the list of 31 NIST keywords, we found that some of them are non-discriminatory. For example, the keyword `validation' is frequently used as a generic term in AI, but it does not serve as a specific keyword for identifying RAI papers. After eliminating these non-discriminatory terms, we retained 18 keywords and included an additional 7 keywords related to `Sustainability,' since it is a significant topic in the public RAI debate.}
    \label{tab:nist-keywords}
\end{table}

\noindent\textbf{Removed keywords}. These keywords were removed from our query because they were non-discriminative as they are often used to describe generally AI. The 19 keywords included: validation, reliability, correctness, accuracy, robustness, generalizability, authenticity, quality, measurability, dependability, capability, safety, security, resilience, confidentiality, integrity, availability, usability, controllability.

\vfill\eject
\section{Analyzed Venues}
\begin{figure}[h!]
    \centering
    \includegraphics[width=0.45\textwidth]{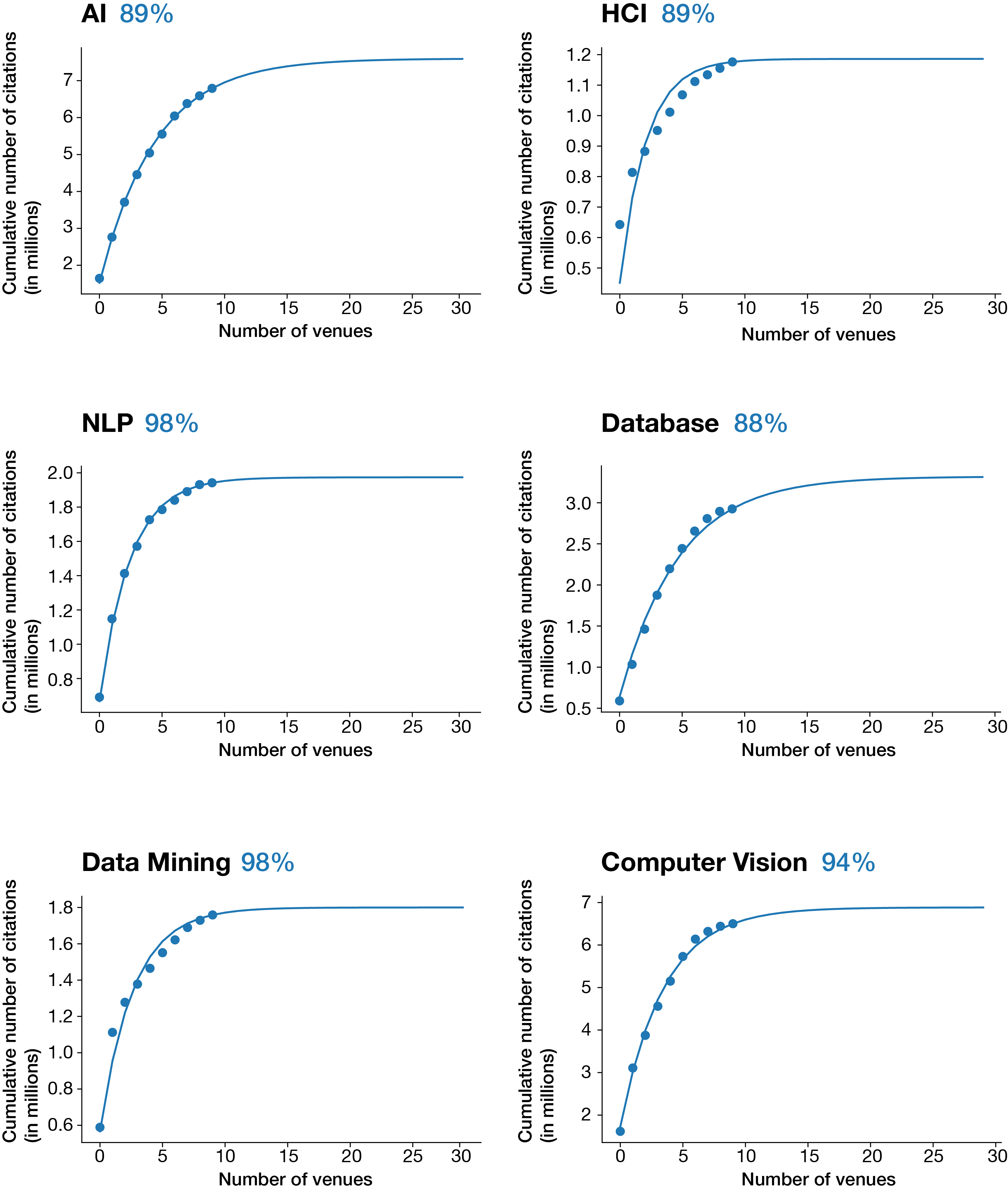}
    \caption{The cumulative number of citations in millions as a function of top venues considered in each research area. The top 10 venues in the six research areas (AI, Database, NLP, HCI, Data Mining, and Computer Vision) cover more than 88\% of the total citations in each area. The numbers at the top of each subplot show the total coverage of citations from the top 10 venues in each area.}
    \label{fig:venues_selection_appendix}
\end{figure}

\clearpage

\begin{table*}[h!]

\small

\resizebox{\textwidth}{!}{
    \tiny
    \begin{tabular}{p{2.5cm}lr}
        \toprule
        \textbf{Category} & \textbf{Venue} & \textbf{\# Papers} \\
        \midrule
        Artificial Intelligence & Neural Information Processing Systems (NeurIPS) & 16049 \\
         & International Conference on Learning Representations (ICLR) & 5679 \\
         & International Conference on Machine Learning (ICML) & 9139 \\
         & AAAI Conference on Artificial Intelligence (AAAI) & 15120 \\
         & Expert Systems with Applications & 17051 \\
         & IEEE Transactions on Neural Networks and Learning Systems & 5419 \\
         & IEEE Transactions On Systems, Man And Cybernetics Part B, Cybernetics & 1917 \\
         & Neurcomputing & 18004 \\
         & International Joint Conference on Artificial Intelligence (IJCAI) & 11721 \\
         & Applied Soft Computing & 8756 \\
        \hline
        Human-Computer Interaction & Computer Human Interaction (CHI) & 10640 \\
         & Proceedings of the ACM on Human-Computer Interaction (CSCW) & 2039 \\
         & Proceedings of the ACM on Interactive, Mobile, Wearable and Ubiquitous Technologies (IMWUT) & 1408 \\
         & IEEE Transactions on Affective Computing & 883 \\
         & International Journal of Human-Computer Studies & 2268 \\
         & Behaviour and Information Technology & 2160 \\
         & ACM/IEEE International Conference on Human Robot Interaction (HRI) & 3475 \\
         & International Journal of Human-Computer Interaction & 2145 \\
         & Virtual Reality & 1279 \\
         & International Conference on Intelligent User Interfaces (IUI) & 1849 \\
        \hline
        Computer Vision & IEEE/CVF Conference on Computer Vision and Pattern Recognition (CVPR) & 9812 \\
         & European Conference on Computer Vision (ECCV) & 5908 \\
         & IEEE/CVF International Conference on Computer Vision (ICCV) & 4775 \\
         & IEEE Transactions on Pattern Analysis and Machine Intelligence (TPAMI) & 7398 \\
         & IEEE Transactions on Image Processing & 8912 \\
         & Pattern Recognition & 9837 \\
         & IEEE/CVF Computer Society Conference on Computer Vision and Pattern Recognition Workshops (CVPRW) & 3684 \\
         & Medical Image Analysis & 2530 \\
         & IEEE/CVF Winter Conference on Applications of Computer Vision (WACV) & 2815 \\
         & International Journal of Computer Vision (IJCV) & 3313 \\
        \hline
        Data Mining \& Analytics & ACM SIGKDD International Conference on Knowledge Discovery \& Data Mining (KDD) & 5195 \\
         & IEEE Transactions on Knowledge and Data Engineering & 8546 \\
         & International Conference on Artificial Intelligence and Statistics (AISTATS) & 3783 \\
         & ACM International Conference on Web Search and Data Mining (WSDM) & 1565 \\
         & Journal of Big Data & 1350 \\
         & Wiley Interdisciplinary Reviews: Data Mining and Knowledge Discovery & 1474 \\
         & IEEE International Conference on Big Data (Big Data) & 6016 \\
         & IEEE International Conference on Data Mining (ICDM) & 1544 \\
         & ACM Conference on Recommender Systems (RecSys) & 1649 \\
         & Knowledge and Information Systems & 3009 \\
        \hline
        Natural Language Processing & Meeting of the Association for Computational Linguistics (ACL) & 9314 \\
         & Conference on Empirical Methods in Natural Language Processing (EMNLP) & 7565 \\
         & Conference of the North American Chapter of the Association for Computational Linguistics (HLT-NAACL) & 3518 \\
         & Transactions of the Association for Computational Linguistics (TACL) & 214 \\
         & International Conference on Computational Linguistics (COLING) & 5793 \\
         & International Conference on Language Resources and Evaluation (LREC) & 7994 \\
         & Workshop on Machine Translation (WMT) & 1330 \\
         & International Workshop on Semantic Evaluation (SemEval) & 2582 \\
         & Conference on Computational Natural Language Learning (CoNLL) & 989 \\
         & Computer Speech and Language & 1254 \\
        \hline
        Database \& Information Systems & International World Wide Web Conferences (WWW) & 7958 \\
         & IEEE Transactions on Knowledge and Data Engineering & 8546 \\
         & ACM SIGIR Conference on Research and Development in Information Retrieval & 5840 \\
         & Information Processing \& Management & 3519 \\
         & ACM International Conference on Information and Knowledge Management (CIKM) & 7106 \\
         & International Conference on Very Large Databases (VLDB) & 6248 \\
         & ACM International Conference on Web Search and Data Mining (WSDM) & 1565 \\
         & ACM SIGMOD International Conference on Management of Data & 4940 \\
         & Journal of Big Data & 1350 \\
         & International Conference on Data Engineering (ICDE) & 5575 \\
         \hline
        Responsible AI & ACM Conference on Fairness, Accountability, and Transparency (FAccT) & 339 \\
         & AAAI/ACM Conference on AI, Ethics, and Society (AIES) & 474 \\
         \bottomrule
    \end{tabular}
}
\caption{Distribution of papers (including non-RAI papers) over the years for each research area. Venues are ranked based on their h5 index impact factor. According to Google Scholar, h5 index is the h-index for articles published in the last five complete years; it is the largest number $h$ such that $h$ articles published in 2018-2022 have at least h citations each.}
\label{table:venues_appendix}
\end{table*}

\clearpage

\end{document}